\documentclass[twocolumn,prb,showpacs,showkeys,nobibnotes,preprintnumbers,floatfix]{revtex4-1}

\usepackage{amsmath}
\usepackage{amssymb}
\usepackage{amsfonts}
\usepackage{graphicx}
\usepackage{dcolumn}
\usepackage{bm}
\usepackage{color}

\newcommand{\eg}{e.\,g.\ }
\newcommand{\ie}{i.\,e.\ }

\newcolumntype{C}{>{\hspace*{\fill}}c<{\hspace*{\fill}}}

\begin{document}

\title{Optimized Wang-Landau sampling of lattice polymers: Ground
  state search and folding thermodynamics of HP model proteins}

\author{Thomas W\"ust}
\affiliation{Swiss Federal Research Institute WSL, Z\"urcherstrasse 111, CH-8903 Birmensdorf, Switzerland}
\email{thomas.wuest@wsl.ch}
\author{David P.\ Landau}
\affiliation{Center for Simulational Physics, The University of Georgia, Athens, GA 30602, USA}


\begin{abstract}
  Coarse-grained (lattice-) models have a long tradition in aiding
  efforts to decipher the physical or biological complexity of
  proteins. Despite the simplicity of these models, however, numerical
  simulations are often computationally very demanding and the quest
  for efficient algorithms is as old as the models
  themselves. Expanding on our previous work [T.\ W\"ust and
    D.\ P.\ Landau, Phys.\ Rev.\ Lett.\ \textbf{102}, 178101 (2009)],
  we present a complete picture of a Monte Carlo method based on
  Wang-Landau sampling in combination with efficient trial moves
  (pull, bond-rebridging and pivot moves) which is particularly suited
  to the study of models such as the hydrophobic-polar (HP) lattice
  model of protein folding. With this generic and fully blind Monte
  Carlo procedure, all currently known putative ground states for the
  most difficult benchmark HP sequences could be found. For most
  sequences we could also determine the entire energy density of
  states and, together with suitably designed structural observables,
  explore the thermodynamics and intricate folding behavior in the
  virtually inaccessible low-temperature regime. We analyze the
  differences between random and protein-like heteropolymers for
  sequence lengths up to 500 residues. Our approach is powerful both
  in terms of robustness and speed, yet flexible and simple enough for
  the study of many related problems in protein folding.
\end{abstract}



\maketitle

\section{\label{sec:intro}Introduction}

Coarse-grained (lattice-) models play an important role in clarifying
questions of generic understanding in protein folding and protein
structure prediction.\cite{dill:protein_sci:99, kolinski:polymer:04}
With the aim of separating relevant features from unimportant details,
such ``minimalist'' models allow us to capture only those forces that
effectively drive a system and thus to eventually find the ``basic
laws'' behind a particular phenomenon. Arguably one of the simplest
protein models is the hydrophobic-polar (HP) lattice model introduced
by Dill et al.\cite{dill:biochemistry:85, *lau:macromolecules:89}
Classifying the 20 amino acids in just two types (hydrophobic and
polar), the HP model contains only two physical ingredients: an
\emph{excluded volume} effect (self-avoiding walk on a lattice) and an
effective monomer-monomer interaction among non-bonded nearest
neighbor H monomers, mimicking the \emph{hydrophobic interaction}
which is considered a main driving force of protein folding. Owing to
the HP and similar models, many fundamental concepts and questions,
\eg the relationship between protein sequence and structure, the
notions of energy landscapes and folding funnels, thermodynamic
transitions towards and stability of the native state, or the kinetic
mechanisms of folding, could be systematically investigated by means
of computer simulations.\cite{sali:nature:94,
  shakhnovich:curr_opin_struct_biol:97, dill:protein_sci:99}

Minimalist protein models also laid a basis for the study of many
related problems of biological interest. Examples include protein
aggregation (multi-chain systems),\cite{cellmer:pnas:05,
  *zhang:biophys_chem:08} surface
adsorption,\cite{castells:phys_rev_E:02, *bachmann:phys_rev_E:06,
  radhakrishna:j_chem_phys:12} protein folding in heterogeneous (\eg
membranes)\cite{gersappe:j_chem_phys:93, *bonaccini:phys_rev_E:99,
  *zhdanov:proteins:01} and crowded or confined
environments,\cite{ping:j_chem_phys:03} or the formation of knots in
proteins.\cite{faisca:phys_biol:10}

A common requirement for such studies is a generic Monte Carlo method
capable of efficiently sampling the conformational spaces of the
system. Despite the formal simplicity and minimalistic framework of
the HP and related models, however, numerical simulations are often
computationally very demanding. The reliable numerical estimation of
the low temperature ($T$) thermodynamic behavior of these models is
particularly challenging. The origin of this difficulty is twofold:
(i) Steric constraints imposed by the underlying lattice (attrition
problem) and other conformational or energy barriers lead to rough
energy landscapes and, thus, hamper the proper sampling of
conformational space. These issues become particularly noticeable at
low $T$, where the polymers exhibit very compact conformations. (ii)
The low-energy conformations often have a very low degeneracy and the
energy density of states (DOS), $g(E)$, (which measures energy
degeneracy) increases rapidly with chain length or number of energy
levels.

Whereas problem (i) is not unique to protein-like models and also
appears in simulations of (off-) lattice
homopolymers,\cite{rampf:j_polym_sci_B_polym_phys:06,
  *seaton:phys_rev_E:10} problem (ii) is a direct consequence of
protein sequence specificity. For example, the HP sequence with 103
monomers (investigated below) has 59 energy levels and $g(E)$ spans
more than 50 orders of magnitude, whereas the corresponding
interacting self-avoiding walk (\ie homopolymer of the same length
with H monomers only) contains 139 energies but with a $g(E)$ range of
38 orders of magnitude only.

Together, these obstacles pose a significant challenge to numerical
studies and have led to the invention of various sophisticated, but
often also specialized algorithms, such as sequential
importance sampling\cite{zhang:j_chem_phys:02} and chain-growth
methods,\cite{otoole:j_chem_phys:92, beutler:protein_sci:96} the most
prominent example being the pruned-enriched Rosenbluth method
(PERM)\cite{grassberger:phys_rev_E:97} and its
variants\cite{frauenkron:phys_rev_lett:98, *bastolla:proteins:98,
  hsu:j_chem_phys:03, *hsu:phys_rev_E:03} and
flat-histogram/multicanonical
versions,\cite{prellberg:phys_rev_lett:04, bachmann:phys_rev_lett:03,
  *bachmann:j_chem_phys:04} equi-energy
sampling,\cite{kou:j_chem_phys:06} multi-self overlap ensemble Monte
Carlo,\cite{iba:j_phys_soc_jpn:98, *chikenji:phys_rev_lett:99}
fragment regrowth Monte Carlo,\cite{zhang:j_chem_phys:07} etc. (see
also references therein).

Since finding the lowest energy conformation(s) for a given HP
sequence is an NP-complete problem,\cite{berger:j_comput_biol:98,
  *crescenzi:j_comput_biol:98} the model has also raised interest in
the computer science community as a challenge in combinatorial
optimization. Targeted specifically to the search of minimal energy
conformations, several (heuristic) algorithms have been proposed,
ranging from genetic algorithms\cite{unger:j_mol_biol:93,
  konig:biosystems:99} and evolutionary Monte
Carlo\cite{liang:j_chem_phys:01} or ant colony
models\cite{shmygelska:bmc_bioinformatics:05} to constraint-based
algorithms.\cite{backofen:constraints:06} However, we stress that such
approaches, while potentially advancing research in protein
\emph{structure prediction}, are only of limited use to the
understanding of the thermodynamics/kinetics of protein
\emph{folding}.

One motivation behind this work was the desire to develop an algorithm
with the power to attack the HP model and the flexibility to be
applicable also to more general setups as described above. Expanding
on previous work,\cite{wuest:comput_phys_commun:08,
  *wuest:phys_rev_lett:09} we present here a generic, fully blind and
fast Monte Carlo procedure, based on the combination of Wang-Landau
sampling\cite{wang:phys_rev_lett:01, *wang:phys_rev_E:01} with
efficient Monte Carlo trial moves, which is very successful both in
finding low energy conformations \emph{and} obtaining thermodynamic
properties for HP-like models over the \emph{entire} temperature
range, including the difficult to access low-temperature
regime. Wang-Landau sampling has been shown to be very efficient and
robust in various fields of statistical physics including simulations
of spin systems,\cite{zhou:phys_rev_lett:06, torbrugge:phys_rev_B:07}
polymers and proteins,\cite{rampf:j_polym_sci_B_polym_phys:06,
  *seaton:phys_rev_E:10, gervais:j_chem_phys:09} or even numerical
integration,\cite{troster:phys_rev_E:05, li:comput_phys_commun:07} (see
also references therein). Instead of sampling a system at a single
temperature, one estimates $g(E)$ from which thermodynamic quantities
at any temperature can be derived. By performing a random walk in
energy space, the algorithm is well suited for overcoming energy
barriers typically encountered in complex free energy
landscapes. However, in order to fully exploit the capabilities of the
algorithm for systems which also have complex conformational spaces,
suitably chosen Monte Carlo trial moves must be introduced. Finding an
optimal interplay between Monte Carlo driver, trial move set and
efficient implementation is a significant achievement of the present
study.

The paper is organized as follows: We first provide a detailed account
of our methodology (Sec.~\ref{sec:method}). Comparing with two other
successful methods, we then demonstrate its efficiency in finding low
energy states for common benchmark HP sequences
(Sec.~\ref{sec:hp:ground}). In Sec.~\ref{sec:hp:thermo} we study the
thermodynamics of these model proteins and show how to elucidate the
subtle conformational changes occurring at low temperature by means of
suitable structural observables. In Sec.~\ref{sec:hp:500mers}, we
investigate the generic differences in the thermodynamic behavior
between random and protein-like heteropolymers for long chain
lengths. Finally, in Sec.~\ref{sec:conclusion} gives our conclusion.

\section{\label{sec:method}Model and Simulation Method}

HP model proteins consist of isolated self-avoiding walks (SAWs) on a
regular lattice with each site of the walk being occupied by a monomer
(either polar or hydrophobic). Here, we only consider the commonly
studied square (2D) and simple cubic (3D) lattices. Self-avoidance
means that no lattice site can be occupied by more than one monomer at
any time. $N$ denotes the number of monomers (\ie the SAW has length
$N-1$). The energy $E$ of a protein conformation is defined by the
number of non-bonded nearest neighbor contacts among hydrophobic
monomers, each of which being associated with an attractive energy
$-\epsilon$.

In the canonical ensemble the partition function is
\begin{equation}
  \label{eq:part_func}
  Z_N(T) = \sum_{\omega \in \Omega_N} e^{-E/kT} = \sum_E g(E) e^{-E/kT},
\end{equation}
where $k$ is Boltzmann's constant and $T$ is the temperature. The
first sum runs over the set of all conformations $\Omega_N$ while the
second sum over all energies $E$ introduces the energy density of
states, $g(E)$. Since $g(E)$ does not depend on $T$, the second form
allows us to calculate $Z_N(T)$ at \emph{any} $T$ and is thus the
target quantity of our interest.

Despite the conceptual simplicity of the model, estimating $g(E)$ over
the entire - and in particular low - energy range of long HP protein
sequences is a non-trivial task (see the Introduction). Principally,
it can be subdivided into three aspects: (A) choice of Monte Carlo
sampling scheme (Monte Carlo driver); (B) choice of appropriate and
efficient Monte Carlo trial moves; (C) efficient and fast
implementation.

\subsection{\label{sec:wls}Wang-Landau sampling}

In Wang-Landau sampling, the \emph{a priori} unknown density of states
$g(E)$ of energy $E$ is iteratively determined by performing a random
walk in energy space seeking to sample a flat energy distribution. The
expression ``random walk in energy space'' emanates from the fact that
conformations are sampled with a probability approaching $\sim 1/g(E)$
resulting in a uniform distribution in $E$. Note that the method can,
in principle, be applied to any type of observable (not only energy)
or to several observables simultaneously.\cite{zhou:phys_rev_lett:06,
  torbrugge:phys_rev_B:07}

Initially, $g(E)=1, \forall E$. A Monte Carlo trial move from a state
(or conformation) $A$ with energy $E_A$ to a state $B$ with energy
$E_B$ is accepted according to the transition probability
\begin{equation}
  \label{eq:wl_transition}
  P(A\rightarrow B)=
  \min\left(1,\frac{g(E_A)}{g(E_B)}\right).
\end{equation}
After each trial move, $g$ is modified as $g(E_t)=f\times
g(E_t)$ ($f>1$ is called the modification factor, initially $f=e^1$)
and, simultaneously, a histogram $H$ is updated, $H(E_t)=H(E_t)+1$,
where $t$ stands for state $B$ if the trial move has been
accepted, otherwise $t\equiv A$. Once the energy distribution in $H$
is sufficiently flat (\ie $H(E)\ge p\langle H(E)\rangle$ for all
energies $E$, where $\langle H(E)\rangle$ is the average histogram and
$0<p<1$ is called the ``flatness criterion''), the modification factor
$f$ is reduced as $f=\sqrt{f}$, $H$ is reset to zero and a new
iteration begins. This process is repeated until $f$ falls below a
threshold ($f_{\text{final}}\approx 1$) at which point $g(E)$ has
converged towards the correct density of states.

The accuracy of the calculated $g(E)$ is controlled by two simulation
parameters: the final modification factor $f_{\text{final}}$ and the
flatness criterion $p$. For a discussion on the choice of these
parameters, see below.

Knowledge of the upper and lower energy boundaries as well as any
nonexistent energy states (energy gaps) within this range, is
essential in the WL algorithm in order to examine the flatness of the
histogram and, ultimately, to control the course of the
simulation. Often, however, the exact energy range is \emph{a priori}
unknown, (hence the use of ground state search algorithms, \eg for the
HP model). To solve this dilemma, performing a \emph{pre-}WL run to
determine the energy range and thereafter fix the sampling to this
range has been proposed.\cite{troster:phys_rev_E:05,
  torbrugge:phys_rev_B:07} The problem with such an approach is
two-fold: (i) The pre-WL run may take a long time to accurately
explore energy space and it is thus unclear when to stop. (ii) Often,
new energy states are found rather late in the course of the
simulation when the running DOS estimate has resulted in a
sufficiently ``flat sampling''.

To overcome these difficulties, the following procedure proved to be
most efficient:\cite{wuest:comput_phys_commun:08,
  *wuest:phys_rev_lett:09} Every time a new energy state
$E_{\text{new}}$ is found, it is marked as ``visited'',
$g(E_{\text{new}})$ is set to $g_{\text{min}}$ (\ie the minimum of $g$
among all previously visited energy states) and the histogram is reset
to zero. The flatness of the histogram is always checked among visited
energy states only, with a sufficiently long interval between
subsequent checks (\eg every $10^6$ MC moves). With this self-adaptive
procedure, new regions of conformation space can be explored while the
current estimate of the DOS is further refined which, in turn, tends
to improved sampling (with respect to a flat histogram in energy
space).

\subsection{\label{sec:mcmoves}Monte Carlo trial moves}

Compared to traditional Monte Carlo schemes, the Wang-Landau algorithm
has been an enormous improvement in overcoming the obstacles typically
encountered near phase transitions. However, the advantageous dynamics
of this Monte Carlo driver is most effective when used in concert with
suitable Monte Carlo trial moves.

Usually, studies of lattice polymers introduce local moves that change
a single bond (end flips), two bonds (corner flips) or three bonds
(crankshafts).\cite{sokal:95} However, such moves suffer from slow
dynamics and are non-ergodic.\cite{madras:j_stat_phys:87} Global
moves, like pivot operations, sample conformational space of
(athermal) self-avoiding walks very
efficiently\cite{madras:j_stat_phys:88} but in the low temperature,
compact, polymer phase their acceptance ratio becomes negligible.

\begin{figure*}

  \includegraphics[width=1.0\textwidth,clip]{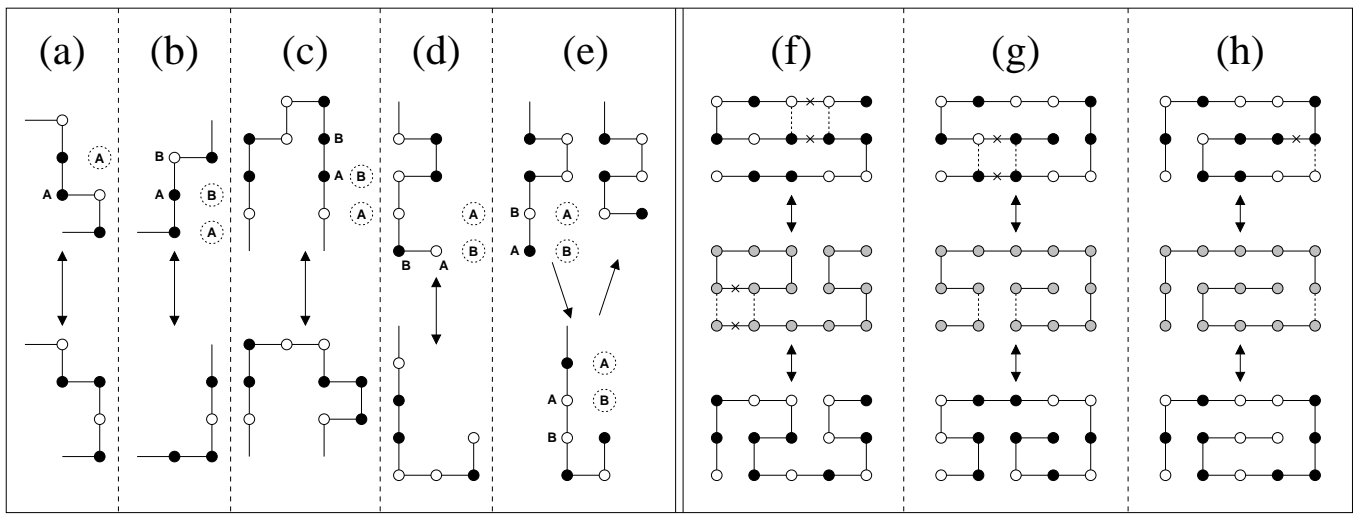}

  \caption{\label{fig:mcmoves}Illustration of MC trial
    moves. \emph{Left panel:} Pull moves: (a) single-bead move (kink
    flip); (b) two-beads move; (c) internal multi-beads move; (d)
    chain-terminal move; (e) chain-terminal move forming a ``hook'';
    this move is not allowed (non-reversible). The dotted circles
    denote the primary (A) and secondary (B) displacement sites of
    monomers A and B, respectively. Subsequent monomers are then
    pulled sequentially to previously occupied sites until the chain
    reaches a new valid configuration. \emph{Right panel:}
    Bond-rebridging moves: (f) chain internal move (``type 1'' in
    Ref.~\onlinecite{deutsch:j_chem_phys:97}) with two consecutive
    cut-and-join steps (in the intermediate stage, the chain is
    divided into a circular and a linear piece); (g) chain internal
    move (``type 2'' in Ref.~\onlinecite{deutsch:j_chem_phys:97}) with
    a single cut-and-join step; (h) chain-terminal move. Bond cutting
    and rejoining are marked with a cross and a dashed line,
    respectively. Note that cutting/rejoining steps may alter the HP
    sequence along the chain (indicated by beads colored in gray in
    the intermediate stages), so once the chain has moved to a new
    conformation, monomer types must be relabeled to ensure the same
    HP sequence. Double arrows indicate reversibility of moves. All
    examples have been adopted from Refs.~\onlinecite{lesh:recomb:03},
    \onlinecite{kou:j_chem_phys:06} and
    \onlinecite{deutsch:j_chem_phys:97}.}

\end{figure*}

The key to our approach is the combination of two ``non-traditional''
Monte Carlo trial moves which complement one another extremely well,
namely pull moves\cite{lesh:recomb:03} and bond-rebridging
moves,\cite{deutsch:j_chem_phys:97} see Fig.~\ref{fig:mcmoves}.

\emph{Pull moves:} This recently introduced trial move allows for the
close-fitting motion of a polymer chain within a confining environment
by continuously ``pulling'' portions of the polymer to unoccupied
neighboring sites of the chain. For a detailed description, see
Ref.~\onlinecite{lesh:recomb:03}. Pull moves are reversible and
ergodic. They feature a good balance between local and global
conformational changes; on average, less than 10 monomers are
displaced during a pull move, keeping the change in energy per move
modest and thus always guaranteeing a reasonable acceptance
ratio. Moreover, pull moves provide an efficient means of folding and
unfolding because of their reptation-like dynamics and the fact that
monomers move largely along paths that were already occupied; hence,
it is less likely that a trial move violates the "excluded volume"
condition. Such features are vital in WL sampling to explore both
conformation and energy space efficiently. Fig.~\ref{fig:mcmoves}
shows some examples of valid pull moves and one invalid
(non-reversible) end pull move (for completeness, we note this case
explicitly here since it was not clearly mentioned in the original
descriptions).

\emph{Bond-rebridging moves:} With decreasing temperature, polymers
form dense and compact conformations leaving only very few unoccupied
lattice sites in the bulk. Therefore, monomer displacing trial moves
are restricted to act on surface beads, and thus become more and more
ineffective with lower $T$ and larger $N$. In contrast,
bond-rebridging (or connectivity altering)
moves\cite{pant:macromolecules:95, ramakrishnan:j_chem_phys:97,
  deutsch:j_chem_phys:97} allow the polymer to change its conformation
even at highest densities by reordering bonds while leaving monomer
positions unchanged. Moreover, they facilitate long range topological
changes, \eg entanglement, which would otherwise require very costly
unfolding/folding processes. This latter feature also becomes
important when the WL sampling of the DOS is split up into energy
subintervals as it substantially reduces the risk of ``locking-out''
conformational space. Fig.~\ref{fig:mcmoves} shows the three types of
bond-rebridging moves applied here; for a detailed discussion, see
Ref.~\onlinecite{deutsch:j_chem_phys:97}.

The combination of bond-rebridging and pull moves yielded an overall
speed-up in WL convergence of up to a factor three as compared to pull
moves only. More importantly, ground states of some HP sequences would
not have been found at all without the concerted interplay of these
two types of trial moves.

To further accelerate global conformational changes, we supplemented
our trial move set with \emph{pivot moves}.\cite{sokal:95} Although
their impact is quite sequence dependent, pivot moves further improved
the sampling (\ie WL convergence time) for most of the HP sequences
considered below. Pivot moves become a prerequisite for chain lengths
$N \gg 100$ since the dynamics of pull moves alone would be too slow
to sample the extended coil conformations of long polymers.

\emph{Detailed balance:} Wang-Landau sampling is a non-Markovian
process and its convergence has been shown without relying on the
condition of detailed balance.\cite{zhou:phys_rev_E:05,
  zhou:phys_rev_E:08} Nonetheless, it is important that trial moves
fulfill detailed balance in order to avoid systematic errors. Unlike,
for instance, single-spin-flip dynamics in the Ising model, where the
number of trial moves is always constant, here the number of valid
trial moves may vary from one conformation to the next. It is
therefore necessary to account for a possible imbalance when
performing a Monte Carlo step from a starting state $A$ to a trial
state $B$. One possibility to do so, is to enumerate all valid trial
moves in $A$ and $B$, and augment the usual Wang-Landau (or
Metropolis-Hastings) acceptance rule with a term compensating for
unequal trial move ratios.\cite{kou:j_chem_phys:06,
  wuest:comput_phys_commun:08, *wuest:phys_rev_lett:09} This
enumeration process is, however, computationally very expensive. A
more efficient Monte Carlo scheme that avoids this counting procedure
in the case of pull moves has been proposed in
Ref.~\onlinecite{swetnam:phys_chem_chem_phys:09} Detailed balance is
guaranteed if a trial move is reversible and the reverse move has the
same probability of being selected as the original move. Therefore, it
is possible to carry out a ``trial-and-error'' procedure where trial
moves are chosen randomly, but with \emph{constant} probability, \ie
independent of the current conformation. Whenever such a trial move is
invalid (\eg resulting in overlapping monomers) the move is rejected
and $g(E), H(E)$ of the old conformation are updated. Otherwise, the
move is accepted according to Eq.~(\ref{eq:wl_transition}). A careful
analysis shows that this procedure can be applied for all three move
types used here. So do all of them fulfill reversibility; sometimes
multiple possibilities exist to go from $A$ to $B$, but this number is
\emph{always} the same in both directions for a given type of
move. Thus, it suffices to employ selection routines for each move
type which ensure that trial moves are chosen with constant
probability:

(i) Pull moves: a list is generated prior to simulation, specifying
the relative displacements of the primary and secondary pull sites
($A$ and $B$ monomers in Fig.~\ref{fig:mcmoves}) for the entirely
elongated polymer chain. This particular conformation features the
maximally possible pull moves,
\begin{equation}
  \label{eq:number_of_pull_moves}
  \begin{split}
    \max n_{\text{pull}} & = (N-2) \times 4(d-1)\\
    & \quad + 2 \times (2d-1+(2d-2)^2),
  \end{split}
\end{equation}
where the first term corresponds to internal pull moves and the second
term to end pull moves, respectively. No other conformation can have
more valid pull moves than this upper bound and all valid pull moves
of any other conformation are contained in this list as a
subset. Then, a Monte Carlo trial step simply consists in choosing a
(valid or invalid) pull move at random from this list.

(ii) Bond-rebridging moves: A random integer $i$ is drawn in the
interval $[0,N]$. If $i=0$ or $1$, a chain terminal rebridging move is
selected (on the one or the other end of the polymer, respectively),
otherwise, a chain internal bond rebridging move (type 1 or 2) is
attempted starting from the bond between monomers $i-1$ and $i$, see
Fig.~\ref{fig:mcmoves}. These trial moves are then carried out as
detailed in Ref.~\onlinecite{deutsch:j_chem_phys:97}.

(iii) Pivot moves: The elements (rotations, reflections and
combinations thereof) of the symmetry group of the $d$-dimensional
lattice $\mathbb{Z}^d$ are orthogonal matrices with integer
entries. There are $d! \times 2^d$ such matrices and they can be
assigned prior to the simulation.\cite{madras:j_stat_phys:88} Thus, to
perform a pivot move trial, a symmetry operation is selected at random
(excluding the identity) and then applied to the shorter part of the
polymer chain subsequent to a randomly set pivot point.

Finally, the ratios among the three types of moves are kept
constant. In all our simulations in this work, we used fractions of
75\%, 23\%, 2\% for pull, bond-rebridging and pivot moves,
respectively. These ratios turned out to provide a good balance to
sample conformational space efficiently over the entire energy range.

By adequately adapting these move fractions, it is possible to achieve
optimal acceptance ratios in any energy range. This can even be
automated by introducing energy dependent (\ie variable) move
fractions. In this case the Wang-Landau acceptance criterion
Eq.~(\ref{eq:wl_transition}) must simply be modified to account for
unequal fractions between the forward and backward (reversible) move
in order to guarantee detailed balance. Although we did not observe a
significant increase in overall efficiency in our WL simulations when
using this technique for the HP sequences considered here, it might
help for very long polymer chains.

\subsection{\label{sec:implement}Efficiency considerations and the choice of $f$ and $p$}

Despite generally higher rejection ratios, the ``trial-and-error''
approach described above outperforms the enumeration
alternative\cite{wuest:phys_rev_lett:09} in speed by one to two orders
of magnitude (depending on chain length and sampled energy range)
without loss in accuracy. To further accelerate our simulations,
additional efficiency mechanisms have been implemented:

\emph{Data structures:} Two different (redundant) data structures have
been used to store a protein conformation. Monomer positions are (i)
stored as $d$-dimensional coordinate vectors and (ii) mapped to the
sites of a lattice (or ``bit table'') of size $L^d$, $L=N+2$, with
periodic boundary conditions. ($L=N+2$ always ensures unhindered
pulling by two lattice units for end pull moves.) Whereas the former
representation facilitates relative monomer displacements or pivot
operations, the latter allows nearest neighbor occupancy queries or
self-avoidance checks in time $O(d)$. Indeed, these very frequent
operations would otherwise scale prohibitively as $O(d \times N)$.

\emph{Energy calculation:} Pull moves usually displace only a small
fraction $\Delta N$ of all monomers; often, this applies to pivot and
bond-rebridging moves too. It is thus more efficient to calculate only
the \emph{change} in energy between subsequent Monte Carlo trial moves
rather than looping over the entire chain to calculate the energy as
long as $\Delta N < N/4$. Together with the usage of a bit table, the
time for an energy calculation scales then as $O(d \times \Delta N)$.

\emph{Random number generator:} For all simulations we used the
Mersenne Twister (MT19937)\cite{matsumoto:tomacs:98} generator because
of its speed. For cross-validation, we used
RANLUX,\cite{luescher:comput_phys_commun:94} a random number generator
of very high statistical strength. Both generators are provided in the
GNU Scientific Library (GSL).\cite{gsl}

\emph{Simulation parameters:} To verify our algorithm and to
understand the influence of the two WL simulation parameters (final
modification factor $f_{\text{final}}$ and flatness criterion $p$) on
the accuracy of the results, we performed extensive tests for two
short HP sequences for which the DOS are known exactly by means of
enumeration, namely a 20mer\cite{unger:j_mol_biol:93} in 2D and a
14mer\cite{bachmann:acta_phys_pol_B:03} in 3D.

\begin{figure}

  \includegraphics[width=1.0\columnwidth,clip]{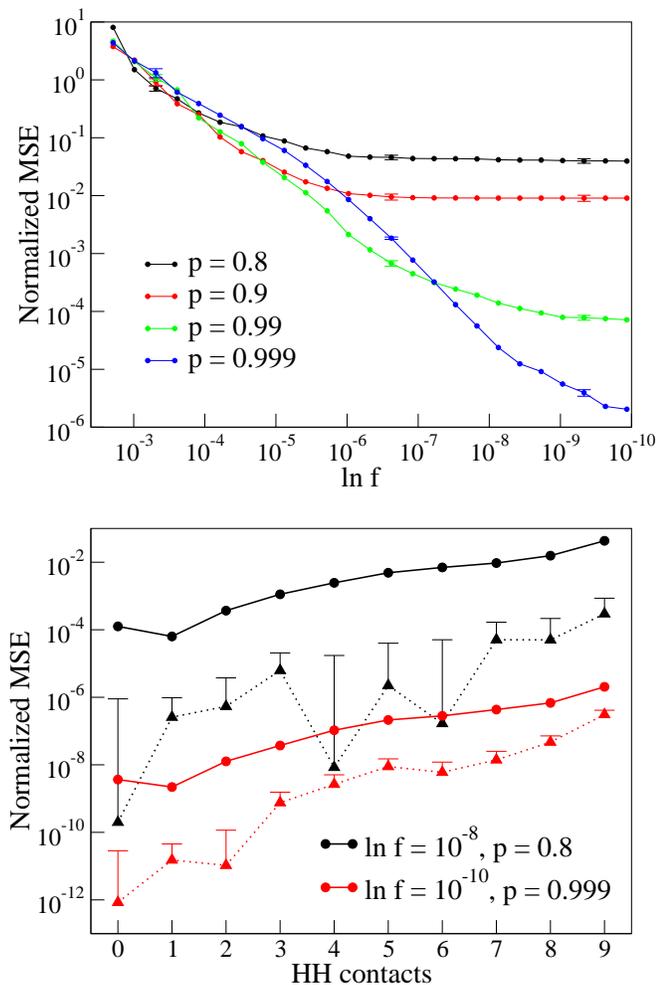}

  \caption{\label{fig:nmse}(Color online) Normalized mean square error
    (NMSE) of DOS estimates for an HP 20mer on the square lattice as a
    function of modification factor $f$ and flatness criterion
    $p$. \emph{Top:} NMSE of $g(E_{\min} = -9)$ vs $\ln f$ for several
    $p$. \emph{Bottom:} NMSE over the entire energy range for two
    pairs of $\ln f_{\text{final}}$ and $p$. The solid lines show the
    entire NMSE (error bars smaller than symbol size) and the dotted
    lines (triangles) show the bias terms only.}

\end{figure}

Fig.~\ref{fig:nmse} shows the normalized mean square error (NMSE) of
our estimates of $g(E)$ for the 20mer and the dependence on $f$ and
$p$ (results for the 14mer are equivalent). The NMSE is defined as
\begin{equation}
  \label{eq:nmse}
  NMSE = \left(\frac{g - \hat{g}}{\hat{g}}\right)^2 + \frac{1}{N-1} \sum_{i=1}^N \left(\frac{g_i - g}{g}\right)^2,
\end{equation}
measuring both the systematic (bias term) and statistical (variance
term) errors. $\hat{g}$, $g$, and $g_i$ stand for the exact, averaged
($N=200$ independent runs have been carried out for statistical
reliability) and individual DOS, respectively, at a certain energy
$E$.

Both the systematic reduction of the NMSE by decreasing $f$ and
increasing $p$ (Fig.~\ref{fig:nmse}, top) as well as the very high
accuracy achieved over the entire energy range for the most stringent
parameter choice ($\ln f_{\text{final}} = 10^{-10}$ and $p = 0.999$,
Fig.~\ref{fig:nmse}, bottom) show clear evidence of the correctness of
our algorithm and in particular of our ``trial-and-error'' approach of
combined usage of pull, bond-rebridging and pivot Monte Carlo trial
moves. Even with $\ln f_{\text{final}} = 10^{-8}$ and $p = 0.8$, our
DOS estimates deviate max.\ a few percent from the exact values
only. (Note that the bias term is generally one to two orders of
magnitude smaller than the variance term.)

Reducing $f$ without increasing $p$ leads to a saturation of the NMSE
while increasing $p$ without sufficiently decreasing $f$ may result in
frozen in systematic errors, see the blue curve in Fig.~\ref{fig:nmse}
(top). These findings on short protein sequences suggest that for $p
\simeq 0.8$ no further gain in accuracy could be achieved when $\ln f
< 10^{-6}$, a result in agreement with WL studies of other polymer
models.\cite{rampf:j_polym_sci_B_polym_phys:06, *seaton:phys_rev_E:10}
However, for longer sequences we observed that the difficult to access
low energy states were often found only during late iterations and
large statistical fluctuations in the low energy DOS estimates still
remained at $\ln f \approx 10^{-6}$. Only further iterations with
reduced $f$ could effectively eliminate such fluctuations. Eventually,
the choice $\ln f_{\text{final}} = 10^{-8}$ and $p = 0.8$ turned out
to yield accurate and reliable DOS estimates over the entire energy
range at feasible computational costs. This choice of parameters
requires three to four orders of magnitude less Monte Carlo trials
until convergence of the DOS as compared to the most rigid pair chosen
here ($\ln f_{\text{final}} = 10^{-10}$ and $p = 0.999$).

\emph{Final note:} The way in which $f$ is decreased affects the
accuracy of the final DOS and the time of convergence of the WL
procedure. In particular, it has been found that, for constant $p$,
decreasing $f$ exponentially (\eg $\sqrt f_{\text{old}} \rightarrow
f_{\text{new}}$) eventually leads to saturation in the error of the
DOS,\cite{belardinelli:phys_rev_E:07, zhou:phys_rev_E:08,
  swetnam:j_comput_chem:11} see also Fig.~\ref{fig:nmse} (top). Some
promising modifications have been proposed for alleviating this
problem, \eg the $1/t$ algorithm\cite{belardinelli:phys_rev_E:07} or
dynamic modification rules for $f$ accounting for the fluctuations in
the histogram $H$.\cite{zhou:phys_rev_E:08, swetnam:j_comput_chem:11}
However, no generic "optimal" rule of reducing $f$ has yet been found
without introducing further (unknown) system-type and -size dependent
simulation parameters which need to be adjusted in order to make the
algorithm effectively converging and outperforming the original
approach. For instance, the normalization factor of Monte Carlo time
in the $1/t$ algorithm is such a parameter and an improper choice may
even lead to non-convergence.\cite{swetnam:j_comput_chem:11} The
process of finding optimal simulation parameters, however, can become
computationally very demanding. On the other hand, our methodological
experience for lattice polymers has shown that a good choice of
efficient MC trial moves has a much larger impact on the overall
performance and efficiency of the WL procedure than such parameter
"tunings". Therefore, in this study we kept the standard WL rule of
reducing $f$ with its proven robustness over a large spectrum of
problems.

\subsection{\label{sec:observables}Calculation of thermodynamic and structural quantities}

Because the energy density of states $g(E)$ does not depend on
temperature $T$, an estimate of $g(E)$ allows us to calculate the
partition function $Z(T)$ (up to a multiplicative constant), see
Eq.~(\ref{eq:part_func}), and its derived thermodynamic quantities at
\emph{any} temperature. For instance, the internal energy $U(T)$ and
specific heat $C(T)$ are obtained as
\begin{equation}
  U(T) = \frac{1}{Z(T)} \sum_E E g(E) e^{-E/kT} \equiv \langle E \rangle
\end{equation}
and
\begin{equation}
  \label{eq:spec_heat}
  C(T) = \frac{\partial U(T)}{\partial T} =
  \frac{\langle E^2 \rangle - \langle E \rangle^2}{kT^2}.
\end{equation}
Besides thermodynamic quantities, structural observables such as the
radius of gyration, etc., are equally important to provide insight
into the conformational changes taking place with varying
temperature. Principally, such structural observables could be sampled
simultaneously during the late iterations of the Wang-Landau
procedure. Often though, some experimentation is required to find
appropriate observables; thus, it may be more convenient to sample
them in a second simulation step.

Here, a ``Wang-Landau resampling'' procedure has been employed, where
the final $g(E)$ obtained from WL sampling is further updated (using a
small modification factor, $\ln f \leq 10^{-7}$) and the same Monte
Carlo acceptance criterion as in Eq.~(\ref{eq:wl_transition}) is
applied. Alongside the simulation, structural quantities are measured
and the sampling stops when all (previously ``visited'') bins of the
energy histogram $H(E)$ have reached a sufficient number of hits (of
the order of $10^8$/$10^9$). The thermodynamic average of a quantity
$Q$ is then obtained from
\begin{equation}
  \label{eq:struct_observ1}
  \langle Q \rangle (T) = \frac{1}{Z(T)} \sum_E e^{-E/kT} g(E) \overline{Q}(E).
\end{equation}
$\overline{Q}(E)$ is the average of $Q$ at energy $E$, \ie
\begin{equation}
  \label{eq:struct_observ2}
  \overline{Q}(E) = \frac{\sum_Q Q H(E,Q)}{\sum_Q H(E,Q)},
\end{equation}
where $H(E,Q)$ is the two-dimensional histogram in $E$ and
$Q$. Because of its advantageous dynamics, this WL resampling
procedure turned out to cover conformational space more uniformly and
faster (including conformational regions of low energy) than with
standard multicanonical sampling.\cite{landau:09, *newman:99} However,
an accurate estimate of $g(E)$ is essential for this procedure to
yield reliable results.

\section{\label{sec:results}Results and Discussion}

\subsection{\label{sec:hp:ground}Ground state search}

Various benchmark HP sequences have been designed, either as
simplified counterparts to real proteins (\eg a 103mer for Cytochrome
C),\cite{lattman:biochemistry:94} or just for the purpose of algorithm
testing.\cite{unger:5th_procs:93, yue:pnas:95b} The longest sequence
explored systematically so far, contains 136
residues.\cite{lattman:biochemistry:94}

First, we applied our method to a series of ten 48mers
(3D),\cite{yue:pnas:95b} which has been used extensively for testing
of algorithmic efficiency, see Table~\ref{tbl:seqs3D48}. For each HP
sequence we measured the lowest energy found ($E_{\min}$), the time
between independent hits of a ground state ($t_{\text{hit}}$, in this
study defined as the time of a ``round trip'', where a conformation
with zero energy must have been visited at least once between two
consecutive hits of a ground state) and the time until convergence of
the density of states over the entire energy interval $[E_{\min},0]$
($t_{\text{DOS}}$). For statistical significance, WLS timings depict
the interquartile mean of 20 independent WL runs.

\begin{table}

  \caption{\label{tbl:seqs3D48}Comparison of performance for different
    methods on a series of ten HP sequences with chain lengths $N =
    48$ in 3D.\cite{yue:pnas:95b} $E_{\min}$ denotes the minimum
    energy reported by all methods. Columns 3-5 give the times (in
    minutes) between independent hits of $E_{\min}$
    ($t_{\text{hit}}$). The last column depicts the WL convergence
    times of the DOS in the energy interval $[E_{\min},0]$ with $\ln
    f_{\text{final}} = 10^{-8}$ and $p = 0.8$. For details of nPERMis
    and FRESS, see Refs.~\onlinecite{hsu:j_chem_phys:03,
      *hsu:phys_rev_E:03} and \onlinecite{zhang:j_chem_phys:07},
    respectively; for nPERMis, we only list the timings for the
    non-biased version.}

  \begin{ruledtabular}
    \begin{tabular}{ccdddd}
      Seq. & $E_{\min}$ &
      \multicolumn{1}{c}{WLS\footnotemark[1]} &
      \multicolumn{1}{c}{nPERMis\footnotemark[2]} &
      \multicolumn{1}{c}{FRESS\footnotemark[3]} &
      \multicolumn{1}{c}{$t_{\text{DOS}}$(WLS)}\\\hline
       1 & -32 & 0.32 &  0.63 & 0.72 & 23.1\\
       2 & -34 & 0.84 &  3.89 & 0.88 & 54.9\\
       3 & -34 & 0.68 &  1.99 & 0.77 & 28.3\\
       4 & -33 & 0.59 & 13.45 & 0.53 & 25.5\\
       5 & -32 & 0.23 &  5.08 & 0.72 & 15.5\\
       6 & -32 & 0.39 &  6.60 & 0.68 & 21.3\\
       7 & -32 & 1.58 &  5.37 & 1.12 & 49.5\\
       8 & -31 & 0.58 &  2.17 & 0.80 & 37.4\\
       9 & -34 & 3.10 & 41.41 & 0.73 & 63.2\\
      10 & -33 & 0.98 &  0.47 & 0.73 & 34.8\\
    \end{tabular}
  \end{ruledtabular}

  \footnotetext[1]{2.8 GHz AMD Opteron 2439}
  \footnotetext[2]{167 MHz Sun ULTRA I}
  \footnotetext[3]{1.4 GHz PC}

\end{table}

Our findings were compared with two other methods for which timing
data are available: (i) an improved variant of the pruned-enriched
Rosenbluth method (nPERMis)\cite{hsu:j_chem_phys:03,
  *hsu:phys_rev_E:03} and (ii) fragment regrowth Monte Carlo via
energy-guided sequential sampling (FRESS).\cite{zhang:j_chem_phys:07}
Although it is clear that a comparison of timings is always limited
because of the different implementations and CPU speeds, it serves,
nonetheless, as a good indicator of performance, in particular,
when considered among several HP sequences.

nPERMis is a thoroughly tested, ``pure'' chain-growth algorithm that
has been very successful on many HP sequences. It is a generic Monte
Carlo sampling scheme able to generate the correct Boltzmann
weights. (Another improved PERM variant,
nPERMh,\cite{huang:phys_rev_E:05} has been exclusively designed for
conformational search and is, therefore, excluded here; for the flat
histogram/multicanonical PERM
versions,\cite{prellberg:phys_rev_lett:04, bachmann:phys_rev_lett:03,
  *bachmann:j_chem_phys:04} only sparse data are available and
unfortunately no timings have been reported.)

In FRESS, Monte Carlo trial moves consist of cutting out and regrowing
pieces of variable length along the protein chain. This update scheme
yields a very efficient conformational search strategy and, so far,
FRESS has been the only method capable of finding putative minimal
energies for all commonly studied benchmark HP sequences. However,
although principally usable for Boltzmann sampling, it has achieved
these results with methodological tuning (``shortcuts'') only, \eg
usage of simulated annealing, neglect of detailed balance, or adoption
of a depth-first-search strategy to regrow chain fragments.

Table~\ref{tbl:seqs3D48} shows that for these still relatively short
sequences, the three methods perform consistently well. They all find
the same putative ground states and the timings between independent
hits are comparable (although the timings for nPERMis fluctuate
considerably more than for WLS and FRESS).

Another series of ten 64mers (3D)\cite{unger:5th_procs:93} showed a
similar picture (details not shown here). We found the same putative
ground state energies as nPERM\cite{grassberger:arxiv:04} and
FRESS. Our timings ranged from a few seconds up to around one hour,
those for nPERM from a few seconds up to 8 hours (on a 2GHz PC); no
timings have been reported for FRESS.

\begin{table}

  \caption{\label{tbl:benchmarks}Comparison of performance for
    different methods on benchmark HP sequences with $N > 50$ on
    square (2D) and simple cubic (3D) lattices. For each sequence, the
    original reference is cited. Columns 3-5 give the times (in hours,
    except otherwise stated) between independent hits of the
    respective energy minima $E_{\min}$ ($t_{\text{hit}}$). The last
    column gives the WL convergence times of the DOS in the energy
    interval $[E_{\min},0]$ with $\ln f_{\text{final}} = 10^{-8}$ and
    $p = 0.8$. For details of nPERMis and FRESS, see
    Refs.~\onlinecite{hsu:j_chem_phys:03, *hsu:phys_rev_E:03} and
    \onlinecite{zhang:j_chem_phys:07}, respectively. NA means no data
    available.}

  \begin{ruledtabular}
    \begin{tabular}{ccdddd}
      Sequence & $E_{\min}$ &
      \multicolumn{1}{c}{WLS\footnotemark[1]} &
      \multicolumn{1}{c}{nPERMis\footnotemark[2]} &
      \multicolumn{1}{c}{FRESS\footnotemark[3]} &
      \multicolumn{1}{c}{$t_{\text{DOS}}$(WLS)}\\\hline
      2D64 \cite{unger:j_mol_biol:93}           & -42 & \multicolumn{1}{c}{$<10$s}       & \multicolumn{1}{c}{$30$}  & <0.33                      &   0.28                    \\
      2D85 \cite{konig:biosystems:99}           & -53 & 0.03                             &  0.17                     & <0.33                      &   1.06                    \\
      2D100a \cite{ramakrishnan:j_chem_phys:97} & -48 & 0.07                             &  0.04                     & <0.33                      &   1.51                    \\
      2D100b \cite{ramakrishnan:j_chem_phys:97} & -50 & 0.17                             &  5.8                      & <0.33                      &   2.49                    \\
      3D58 \cite{dill:pnas:93}                  & -44 & 0.12                             &  0.19                     &  0.09                      &   2.23                    \\
      3D64 \cite{yue:pnas:95a}                  & -56 & 0.09                             &  0.45                     &  0.53                      &   2.61                    \\
      3D67 \cite{yue:pnas:95a}                  & -56 & 0.99                             &  1.12                     &  1.41                      &  13.80                    \\
      3D88 \cite{beutler:protein_sci:96}        & -69 & 0.45                             & \multicolumn{1}{c}{NA}    &                            &   8.30                    \\
                                                & -72 & 8.92                             &                           &  5.03                      & 122.86                    \\
      3D103 \cite{lattman:biochemistry:94}      & -54 & 0.01                             &  3.12                     &                            &   0.65                    \\
                                                & -57 & 0.93                             &                           &  4.47                      &   9.79                    \\
                                                & -58 & \multicolumn{1}{c}{NA}           &                           &                            & \multicolumn{1}{c}{NA}    \\
      3D124 \cite{lattman:biochemistry:94}      & -71 & 0.06                             & 12.3                      &                            &   2.28                    \\
                                                & -75 & \multicolumn{1}{c}{$\approx 104$} &                           & \multicolumn{1}{c}{$<14$d} & \multicolumn{1}{c}{$\ge 10$d}\\
      3D136 \cite{lattman:biochemistry:94}      & -80 & 2.98                             & \multicolumn{1}{c}{$110$} &                            &  34.17                    \\
                                                & -82 & \multicolumn{1}{c}{$\approx 89$} &                           &  6.42                      & \multicolumn{1}{c}{$\ge 10$d}\\
                                                & -83 & \multicolumn{1}{c}{$\ge 10$d}    &                           & \multicolumn{1}{c}{$<14$d} & \multicolumn{1}{c}{NA}    \\
    \end{tabular}
  \end{ruledtabular}

  \footnotetext[1]{2.8 GHz AMD Opteron 2439}
  \footnotetext[2]{667 MHz DEC ALPHA 21264 (seq. 2D85: 167 MHz Sun ULTRA I)}
  \footnotetext[3]{1.4 GHz PC}




\end{table}

As a more stringent test we have selected particular benchmark HP
sequences with $N > 50$ in two and three dimensions (for a listing of
these sequences, see \eg
Ref.~\onlinecite{zhang:j_chem_phys:07}). Table~\ref{tbl:benchmarks}
summarizes the results (WLS timings were obtained as above).
\footnote{WSL timings for the energies -75 (3D124) and -82, -83
  (3D136) have been obtained from WL simulations using already initial
  DOS estimates from preceding runs which effectively found
  these energies.}
Overall,
the performance of an algorithm is now more sequence dependent but two
tendencies are most striking: In 2D, nPERMis shows large timing
differences dependent on the HP sequence whereas WLS and FRESS perform
much more homogeneously. For longer sequences in 3D ($N \ge 88$),
nPERMis performs considerably worse than the other two methods, both
in terms of energy minima found as well as in corresponding
timings. Difficulties in finding the ground state can sometimes be
traced back to certain characteristics of the native structure, \eg
the formation of a very distinct hydrophobic core (3D88) or chain
terminals deeply buried in the interior of the structure
(2D64);\cite{hsu:j_chem_phys:03, *hsu:phys_rev_E:03} see the examples
of ground state structures for each sequence of
Table~\ref{tbl:benchmarks} in Figs.~\ref{fig:seqs2D},
\ref{fig:seq3D67} and \ref{fig:seqs3D}.

\begin{figure}

  \includegraphics[width=0.9\columnwidth,clip]{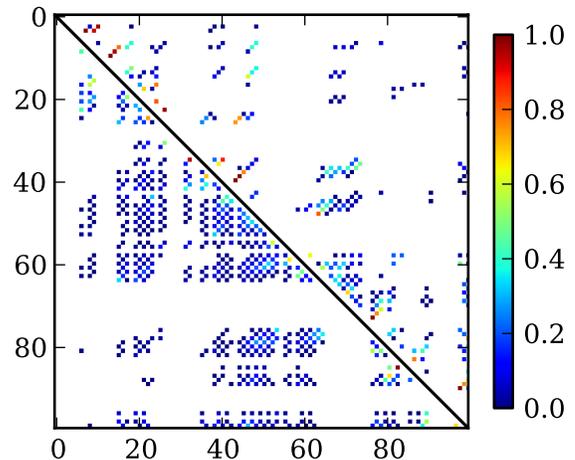}

  \caption{\label{fig:contactmap}(Color online) Contact map density of
    ground states for the sequence 2D100a (lower triangle, $E_{\min} =
    -48$) and 2D100b (upper triangle $E_{\min} = -50$),
    respectively. For each sequence, densities have been calculated
    from a sample of more than 10,000 contact maps of ground states
    contributed from 20 independent WL production runs. Only
    non-vanishing HH contact densities are shown.}

\end{figure}

Often though, the causes are less apparent and looking at a few single
ground state structures is insufficient. Instead, it is necessary to
consider the entire ensemble. For instance, the native states for the
two 100mers in 2D (2D100a and 2D100b, see Fig.~\ref{fig:seqs2D}), look
very similar (as are their minimal energies) and do not feature any
peculiarities. Nonetheless, nPERMis shows timing differences of two
orders of magnitude between the two sequences whereas WLS and FRESS
perform equally well in both cases. In order to better understand this
behavior, we have calculated the contact map densities of the
ensembles of respective ground states for both sequences, see
Fig.~\ref{fig:contactmap}. The figure illustrates clearly that native
structures in case of sequence 2D100a are rather homogeneously
distributed in conformational space whereas for sequence 2D100b they
are concentrated in fewer but denser locations manifesting a lower
degeneracy. As the timings in Table~\ref{tbl:benchmarks} indicate,
this difference does not seem to impact the sampling abilities of
FRESS and WLS. Owing to their efficient Monte Carlo update schemes,
both algorithms can easily traverse between distant regions of
conformational space.

In contrast, nPERMis needs much more time to explore different
portions of conformational space, once growth proceeds far in one
direction (despite sophisticated pruning/enrichment mechanisms). This
problem becomes more severe for longer sequences and denser sampled
conformational spaces. Therefore, we conjecture that this is also the
main cause of troubles for the longer sequences in 3D. Even though
stronger heuristics\cite{beutler:protein_sci:96, hsu:j_chem_phys:03,
  *hsu:phys_rev_E:03, huang:phys_rev_E:05} or
flat-histogram/multicanonical
approaches\cite{prellberg:phys_rev_lett:04, bachmann:phys_rev_lett:03,
  *bachmann:j_chem_phys:04} may partially alleviate these
difficulties, the problem remains inherent in any static chain-growth
algorithm (\ie one that always grows from the same starting point)
when sampling the generally dense conformational spaces at low
temperature. The apparent tendencies exposed by this comparison of
methods in Table~\ref{tbl:benchmarks} clearly contradict the
statements in Ref.~\onlinecite{bachmann:phys_rev_lett:03,
  *bachmann:j_chem_phys:04} where it has been argued that chain-growth
algorithms perform better than Monte Carlo algorithms employing trial
moves in simulating lattice proteins at low temperatures. This
contradiction stems from the fact that the latter methods, such as
Wang-Landau sampling, strongly depend on the chosen trial move set; a
point likely not given enough attention in
Ref.~\onlinecite{bachmann:phys_rev_lett:03, *bachmann:j_chem_phys:04}.

With our strategy, we were able to find \emph{all} currently known
putative ground state energies for these benchmark HP sequences,
inclusive a conformation with energy $-58$ for the sequence 3D103.
This result has also been obtained by constrained-based protein
structure prediction (CPSP).\cite{backofen:constraints:06} However,
CPSP is \emph{not} a ``blind'' search algorithm but rather based on
the threading of a sequence through a set of pre-calculated compact
hydrophobic cores. Thus, it can also not yield thermodynamic
properties and is not applicable to two-dimensional systems where
native structures do not necessarily form compact H-cores. To our
knowledge, our procedure is the only \emph{generic} and fully blind
Monte Carlo sampling scheme achieving these results so far.

Sequence 3D103 demonstrates particularly well the algorithmic
improvements (not just increase of CPU speed) that have been achieved
in the course of time. In 1994, Lattman et
al.\cite{lattman:biochemistry:94} proposed a first ground state energy
$E = -44$ by means of hydrophobic zipper. The sequence then served as
a testing ground for various methods, ranging from contact
interactions\cite{toma:protein_sci:96} ($E = -49$), PERM
variants\cite{hsu:j_chem_phys:03, *hsu:phys_rev_E:03,
  bachmann:phys_rev_lett:03} ($E = -54, -55, -56$),
FRESS\cite{zhang:j_chem_phys:07} ($E = -57$) to
WLS\cite{wuest:phys_rev_lett:09} ($E = -58$). Due to insufficient
sampling of the ground state we could not obtain an estimate of the
relative DOS between the ground and first excited states, but we
certainly now know that the probability of finding a conformation with
energy $E = -58$ is definitely more than 20 orders of magnitude
smaller than for a conformation with $E = -44$.

\subsection{\label{sec:hp:thermo}Thermodynamic properties}

\begin{figure*}

  \includegraphics[width=1.0\textwidth,clip]{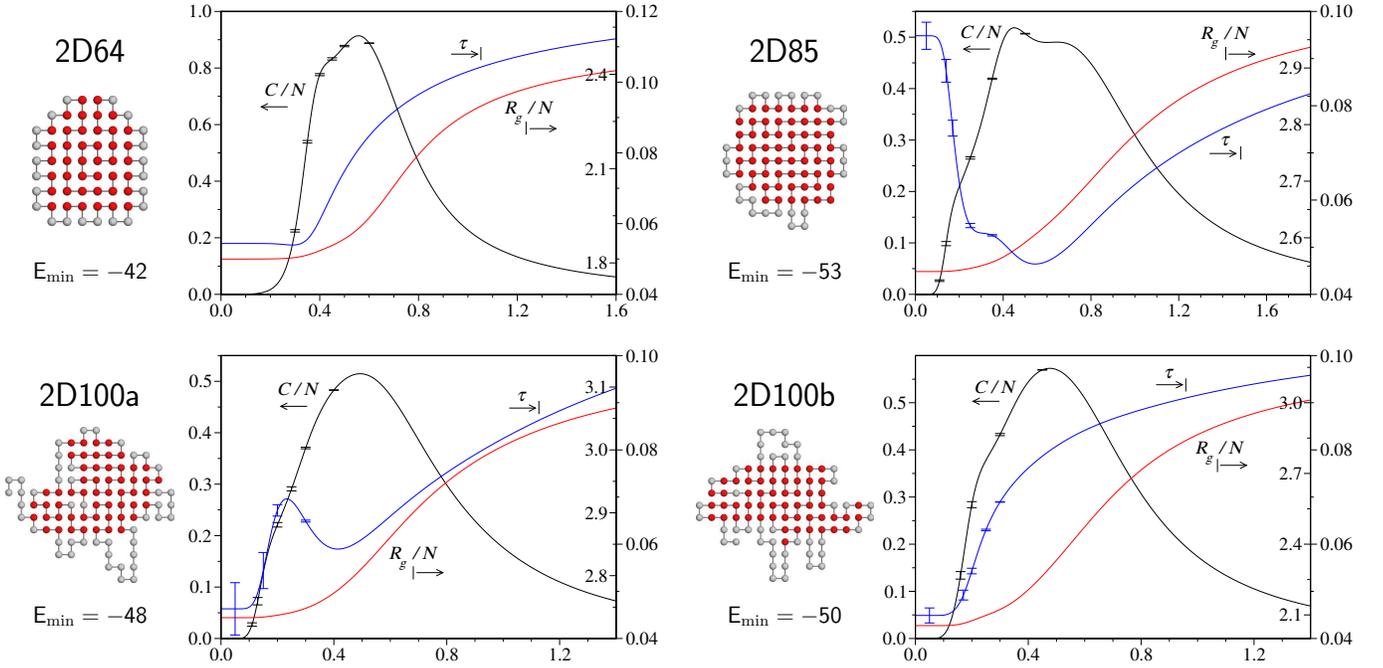}

  \caption{\label{fig:seqs2D}(Color online) Specific heat, $C / N$
    (black curves, left ordinates), and canonical ensemble averages of
    the root mean squared radius of gyration, $R_g / N$ (red curves,
    outer right ordinates), and tortuosity, $\tau$ (blue curves, inner
    right ordinates), as a function of temperature (abscissas: $kT /
    \epsilon$) for several benchmark HP sequences on the square
    lattice (2D). For each sequence, $C$ has been obtained from the
    average DOS of 20 independent WL simulations in the corresponding
    energy interval $[E_{\min}, 0]$, and $R_g$, $\tau$ from 20
    independent WL resampling production runs. Errors have been
    estimated by means of a bootstrap analysis\cite{landau:09,
      *newman:99} with 200 resamples and are shown where visible
    only. For each sequence, a typical ground state conformation is
    shown on the left with red and light-gray beads indicating
    hydrophobic (H) and polar (P) residues, respectively. For further
    explanations, see text.}

\end{figure*}

With regard to conformational search, the robustness ($E_{\min}$'s
found) and performance (timings) of FRESS and our approach are very
much alike with some sequence-dependent differences. The main
advantage of our approach over FRESS, however, lies in the usage of a
simpler, but equally efficient, Monte Carlo trial move set which can
be readily combined with Wang-Landau sampling. This strategy allowed
us not only to find putative ground states but, at the same time, to
obtain estimates of the DOS with very high accuracy ($\ln
f_{\text{final}} = 10^{-8}$ and $p = 0.8$). The DOS gives access to
thermodynamic quantities over the whole temperature range and, in
particular, enables us to better explore the intricate folding
behavior of these model proteins at low temperatures.

\begin{figure}


  \includegraphics[width=1.0\columnwidth,clip]{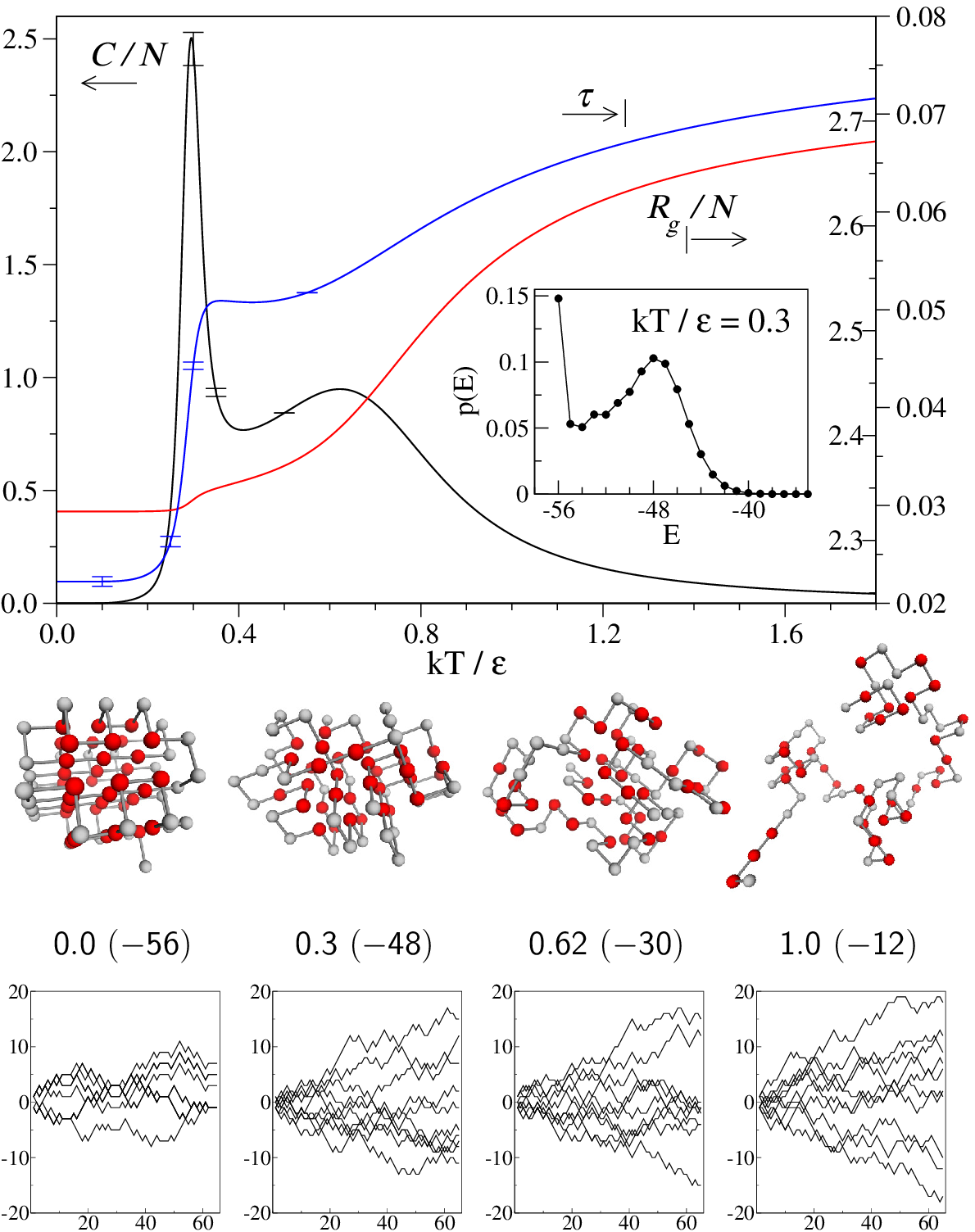}

  \caption{\label{fig:seq3D67}(Color online) Main figure: Same as
    Fig.~\ref{fig:seqs2D} for an HP 67mer on the simple cubic lattice
    (3D).\cite{yue:pnas:95a} The inset figure shows the canonical energy
    distribution, $p(E)$, at the temperature where $C$ takes its
    maximum in the sharp peak. Below, typical structures are shown at
    indicated temperatures and energies ($kT / \epsilon$, $E$). The
    native state of this sequence ($E_{\min} = -56$) resembles an
    $\alpha$/$\beta$-barrel in real proteins. The ground state and the
    structure with $E = -48$ correspond to the two maxima of the
    bimodal distribution of $p(E)$ in the inset figure illustrating
    the significant conformational rearrangements taking place at the
    folding transition. Each of the small figures at
    the bottom shows the ``winding walks'' [$s_i$ vs $i$, see
      Eq.~(\ref{eq:tortuosity})] of 10 structures at corresponding
    energies. For further explanations, see text.}

\end{figure}

\begin{figure*}

  \includegraphics[width=1.0\textwidth,clip]{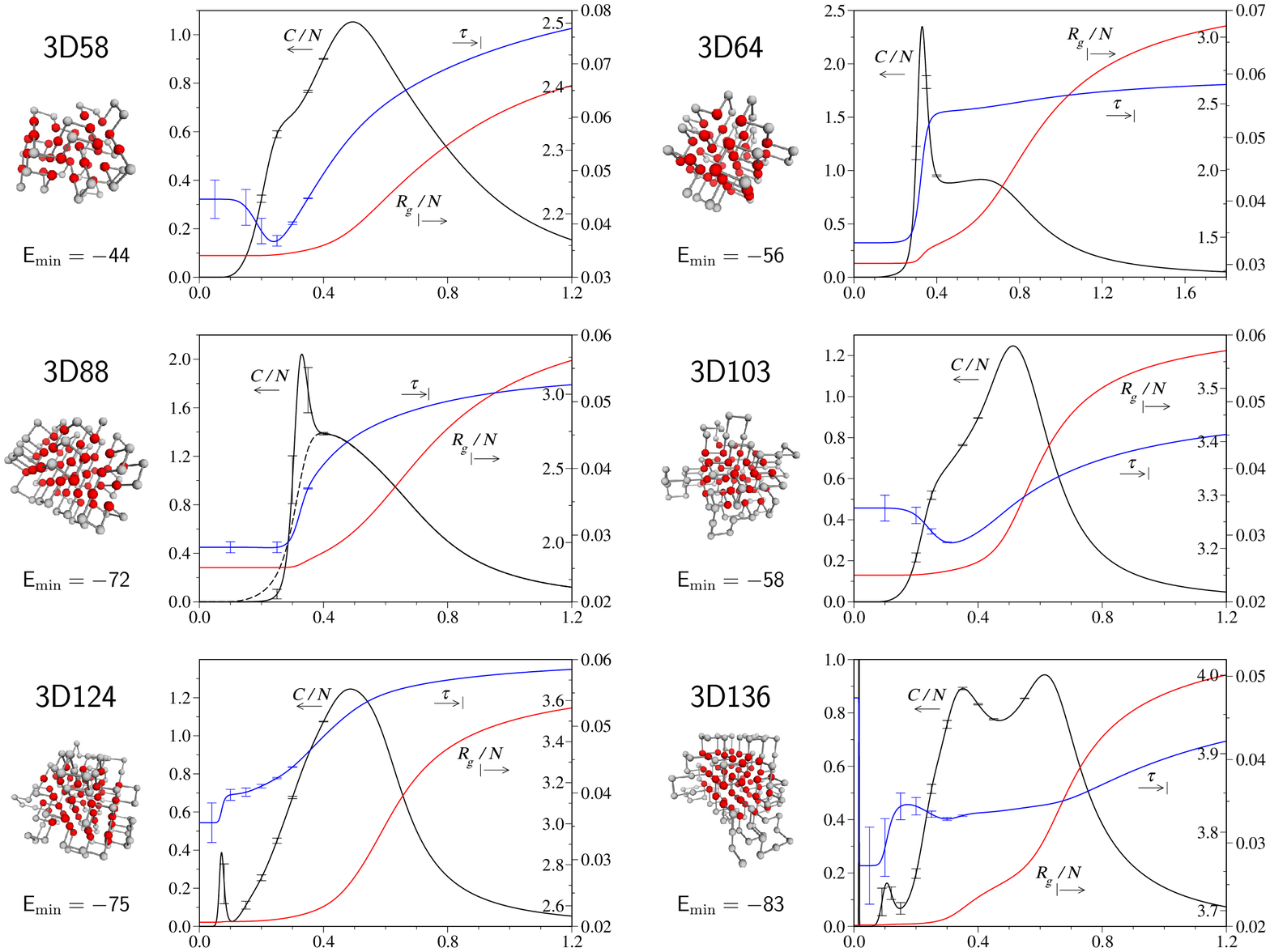}

  \caption{\label{fig:seqs3D}(Color online) Same as
    Fig.~\ref{fig:seqs2D} for several benchmark HP sequences on the
    simple cubic lattice (3D). For sequence 3D88, the dashed line
    shows the specific heat ($C / N$) calculated from the DOS with
    energy range $[-69, 0]$ only. For sequences 3D103 and 3D136, the
    DOS/observables have been only obtained for the energy ranges
    $[-57, 0]$ and $[-82, 0]$, respectively, because of the difficulty
    to sample the respective ground states. Since $\tau$ is very
    sensitive at low $T$, for sequences with $N > 100$, structural
    observables have been averaged from 50 independent WL production
    runs.}

\end{figure*}

Figs.~\ref{fig:seqs2D}, \ref{fig:seq3D67} and \ref{fig:seqs3D} show
the specific heats, $C$ (see Eq.~\ref{eq:spec_heat}), as a function of
temperature ($T$) for the HP sequences of
Table~\ref{tbl:benchmarks}. In free space (as studied here), the $C$
curves of most sequences exhibit two major transitions:

(i) At high $T$, a wide peak in $C$ indicates the collapse of the
protein from a swollen coil to a compact globular (pseudo-) phase; see
also the snapshots of structures at $kT / \epsilon = 1.0$ and $kT /
\epsilon = 0.62$ in Fig.~\ref{fig:seq3D67}. (Note that the term
``phase'' is actually not strictly defined as we are considering
systems of finite size only.) Since the structures in both phases are
still of a random nature, the location of this transition ($T_c$ of
specific heat maximum) shows little sequence dependency (at least for
these still relative short protein-like sequences, see also
Sec.~\ref{sec:hp:500mers}). It depends, however, on the fraction of H
monomers in the sequence. This explains, for instance, the very
similar transition temperatures of sequences 2D100a ($T_c \approx
0.49$, 55\% H) and 2D100b ($T_c \approx 0.48$, 56\% H) and the higher
one for sequence 2D85 ($T_c \approx 0.65$, 69.4\% H). A higher
fraction of H residues raises the collapse transition temperature
because of the generally larger number of HH contacts which need to be
broken upon going from a globular to a coil structure; see also the
discussion in Ref.~\onlinecite{kou:j_chem_phys:06}.

(ii) At low $T$, the specific heat signals another transition which
manifests the folding of the protein from a random globule to its
native state(s). Shape, magnitude and location of this transition are
all very sequence dependent; the curves of $C$ range from a barely
visible shoulder (2D100a, Fig.~\ref{fig:seqs2D}) to a very sharp peak
with a clear first-order-like bimodal energy distribution (3D67, see
also the inset figure and the typical structures at the lower two
temperatures in Fig.~\ref{fig:seq3D67}). Low ground state degeneracy
or energy gaps may give some indication about the nature of this
transition. But ultimately, it is only the entire low energy range of
the DOS which determines the exact character of this
transition. Therefore, accurate sampling of the low energy range is
compulsory to gain a conclusive picture of the low temperature folding
behavior. Otherwise, important features of some sequences do not show
up at all (\eg the sharp peak in the specific heat of sequence 3D88 in
Fig.~\ref{fig:seqs3D}, indicating a pronounced folding transition
which leaves the coil-globule transition as only a flat shoulder
remnant).

None of the HP sequences studied here has an energy gap between ground
and excited states; such a gap has been considered a prerequisite of a
good folder (\ie having a pronounced global energy
minimum).\cite{sali:nature:94} However, the longest two sequences show
very sharp drops (several orders of magnitude) in the DOS between
first excited and ground state (3D124) or between the lowest excited
states (3D136). These ``kinks'' in the DOS cause the sharp, low $T$
specific heat peaks. For sequence 3D124, this is the only signal in
the low temperature regime (no folding transition). The folding
transition of sequence 3D136 is located near $kT / \epsilon \approx
0.35$; below this $T$, two excitation transitions occur (a weak and a
very sharp near $T \rightarrow 0$). Clearly, the locations and
amplitudes of these very low $T$ peaks in $C$ are somewhat uncertain.

Beside the specific heat, structural quantities have been calculated
to get further insight into the folding process. For instance, the
root mean squared radius of gyration, $R_g$,
\begin{equation}
  \label{eq:rad_gyr}
  R_g = \left(\frac{1}{N} \sum_i^N (\bm{r}_i - \bm{r}_{cm})^2\right)^{1/2},
\end{equation}
($\bm{r}_i$ is the position vector of the $i^{\text{th}}$ monomer and
$\bm{r}_{cm}$ is the center of mass), is known to be a well suited
observable to signal the collapse of homo- and
heteropolymers.\cite{bachmann:j_chem_phys:04,
  rampf:j_polym_sci_B_polym_phys:06, *seaton:phys_rev_E:10} Our
figures (red curves in Figs.~\ref{fig:seqs2D}, \ref{fig:seq3D67} and
\ref{fig:seqs3D}) confirm this observation and show a clear
coincidence between the upper maximum in $C$ and the strong decay of
$R_g$. Due to finite size effects,\cite{sharma:j_chem_phys:08} the
steepest slope of $R_g$ is slightly shifted towards higher $T$ as
compared to the specific heat maximum ($T_c$). On the other hand,
$R_g$ shows little or no signal in the low temperature regime. The
polymer end-to-end distance, $R_e$ (not shown here), has very similar
properties as $R_g$ and shows the same S-shaped curve indicating the
collapse. At low $T$, $R_e$ may have a minimum and rise again as $T
\rightarrow 0$. This effect is more pronounced for HP sequences with P
terminals signaling the tendency to push P monomers towards the
surface in order to maximize the formation of a compact hydrophobic
core.\cite{bachmann:j_chem_phys:04} However, measuring the spatial
extent of a polymer only, $R_g$, $R_e$, or similar properties like \eg
box size or surface P number,\cite{kou:j_chem_phys:06} are insensitive
to the sequence specific internal conformational (and topological)
changes taking place upon folding from the globular denatured states
to the ground state.

Measures of conformational ``distance'' to the native state, based on
an adjacency matrix\cite{chan:j_chem_phys:94} or the number of native
contacts (\eg the Jaccard index),\cite{jaccard:01,
  wuest:phys_rev_lett:09} may better serve as a reaction
coordinate. However, for sufficiently long chains, calculations of
such measures can become computationally intractable and they may also
bear some ambiguity (\eg dependence on the choice of move set or large
degeneracy of the native state). Moreover, a requirement of these
observables is that the native state(s) being effectively known to
give meaningful results. Thus, it still remains a challenge to
``design'' a (computationally affordable) structural observable which
reflects the cooperative rearrangements and activity indicated by the
peaks or shoulders in the specific heats at low $T$.

Based on the ideas of the ``DNA walk'',\cite{buldyrev:dna_walk:94} we
propose here a scalar structural observable which might provide better
insight into the \emph{internal} topological changes taking place upon
folding. We define the tortuosity (or writhing/winding measure) of the
protein as
\begin{gather}
  \label{eq:tortuosity}
  \tau = \left( \frac{1}{N-2} \sum_{i=1}^{N-2} (s_i - \bar{s})^2 \right)^{1/2},\\
  \intertext{where}
  s_i = \sum_{j=1}^{i} \bm{r}_{j,j+1} \times \bm{r}_{j,j+2}, \quad 1 \leq i \leq N - 2.
\end{gather}
$\bm{r}_{j,j+1}$ and $\bm{r}_{j,j+2}$ denote the two-dimensional
vectors between monomers $(j, j+1)$ and $(j, j+2)$, respectively (in
the plane spanned by these three monomers). Their cross product
determines whether two consecutive bonds define a left ($-1$) or right
($+1$) turn or are collinear ($0$).\cite{cormen:09} $s_i$ is the
running sum of bond turns along the polymer chain (``winding walk'');
examples of $s_i$ for some conformations at different energies are
shown at the bottom of Fig.~\ref{fig:seq3D67}. The standard deviation
of $s_i$, here denoted as $\tau$, is thus a measure of polymer
tortuosity. (Note that $\tau$ is invariant to the direction of the
running sum $s_i$ along the chain.)

The blue curves in Figs.~\ref{fig:seqs2D}, \ref{fig:seq3D67} and
\ref{fig:seqs3D} display $\tau$ as a function of $T$. Overall, they
reveal the following picture: By lowering $T$, the associated
contraction of the protein results in a reduction of conformational
degrees of freedom and, thus, in a decrease of ``winding freedom'' as
well. At a certain point in the low temperature regime, $\tau$
undergoes a significant change, showing a sharp drop-off (increase) or
passing over a rounded peak (through). The diversity of signals is a
consequence of the different sequences of H and P residues as well as
the different chain lengths. But the important observation is that the
temperature region of this change clearly does \emph{not} coincide
with the collapse transition (where the protein's shape is still
random) but with the low temperature signal in the specific heat (\ie
the folding transition). At this temperature, internal structural
rearrangements take place (\eg breaking of HH contacts of compact
denatured states) which eventually allow the protein to assume its
native state. Being sensitive to exactly such topological
transformations, $\tau$ reflects this folding behavior. It serves,
thus, as a complementary structural quantity to better interpret the
thermodynamic activity observed in the specific heats. It might be
conjectured that the shape of $\tau$ could even give some deeper
insight into the folding characteristics, such as a sharp drop of
$\tau$ indicating a folding funnel or a peak signaling a folding
barrier. Furthermore, the standard deviation is just one (simple)
means of analyzing the ``winding walks'' $s_i$ and other, more
appropriate, measures could be conceived. These questions are left for
further research.

\subsection{\label{sec:hp:500mers} Random vs protein-like HP heteropolymers}

Many of the above HP sequences (Figs.~\ref{fig:seqs2D},
\ref{fig:seq3D67} and \ref{fig:seqs3D}) have been derived from real
proteins and mimic protein-like behavior exhibiting some kind of
collapse and folding transitions. However, for these relatively short
chains ($N \leq 100$), the characteristics and strengths of the two
transitions are highly sequence dependent and subject to strong
surface effects in the folded state (for a perfect cube of $N = 125$,
only about 20\% of the monomers are in the bulk). Although there is no
thermodynamic limit in the exact statistical sense for any single
heteropolymer, here we are interested in identifying the generic
differences between random and protein-like HP sequences in the
long-chain limit.

\begin{figure}

  \includegraphics[width=0.9\columnwidth,clip]{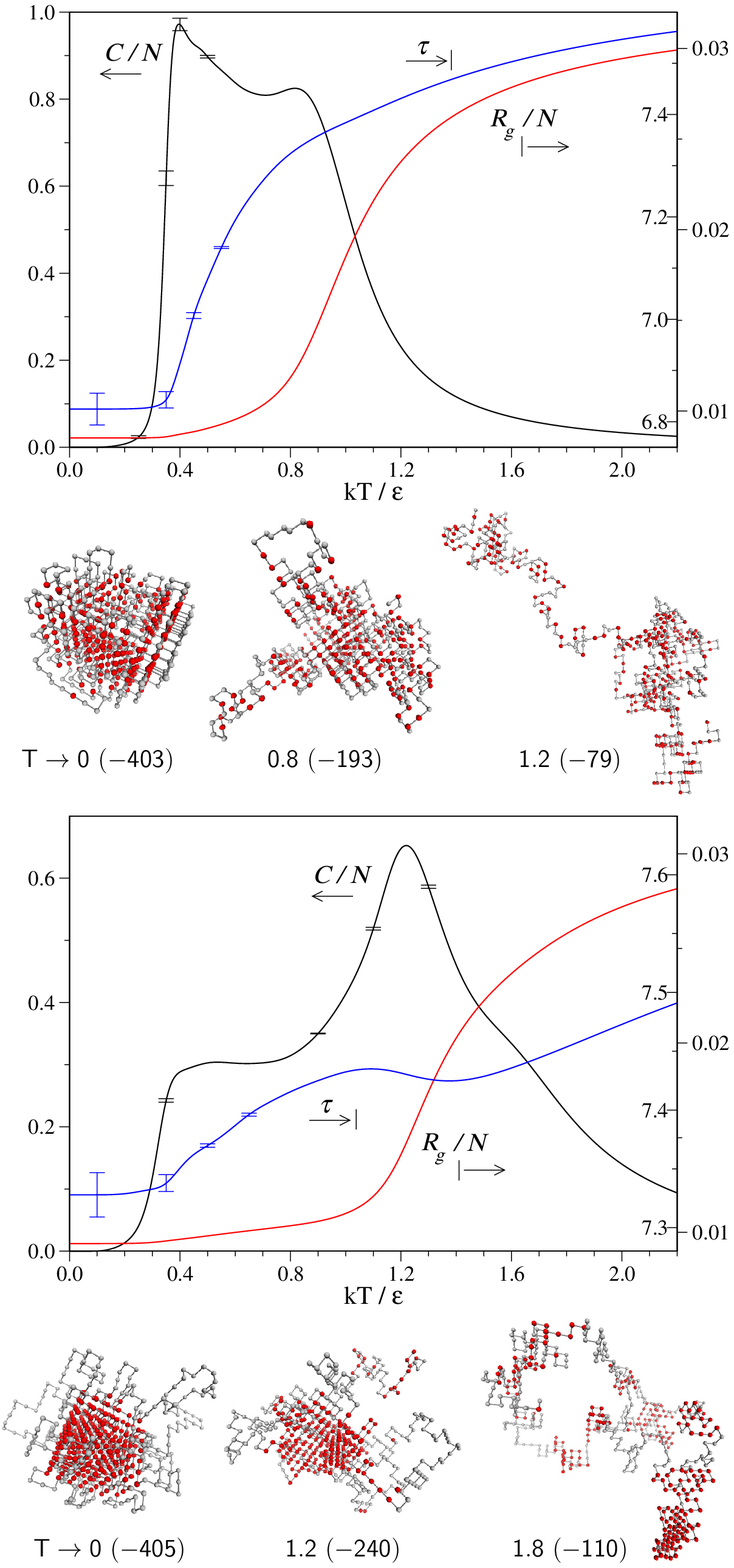}

  \caption{\label{fig:500mers}(Color online) Comparison of
    thermodynamic ($C$) and structural ($R_g, \tau$) properties
    between a random (\emph{top}) and protein-like (\emph{bottom}) HP
    heteropolymer. Both sequences constitute 500 monomers with a 50\%
    : 50\% ratio among H and P. For both sequences, typical structures
    are shown at indicated temperatures and energies ($kT / \epsilon$,
    $E$).}

\end{figure}

To address this question, we have studied several long ($N = 500$) HP
chains, each with 50\% H and 50\% P monomers in either random or
protein-like sequence. The latter have been constructed as
follows:\cite{boelinger:08} Starting from a compact, globular
homopolymer conformation (\ie only H), the 50\% of monomers which are
farthest from the center of mass are marked as P (polar). Repeating
this process for different homopolymer conformations yields
protein-like HP sequences with distinct hydrophobic cores. Longer HP
polymers allowed us also to further test the efficiency of our
procedure. By splitting up the calculation of the DOS into two unequal
energy windows (the lower one covering only 10-15\% of the entire
energy interval with sufficient overlap to the larger window by \eg 25
energy levels), we could accurately estimate the DOS over more than
95\% of the entire energy range of these sequences within a couple of
CPU days. Note that for these chain lengths, finding the ground state
is currently still out of reach for any method due to the
exponentially increasing computational complexity; but this is not a
requirement here.

An important dissimilarity between random and protein-like sequences
becomes already visible in the DOS: For given ratios of H and P
monomers, both types of sequences span roughly the same energy range
(here $E_{\min} \approx -400$) but the DOS is steeper at low energies
for protein-like sequences. This behavior is in agreement with the
notion of folding funnel\cite{dill:protein_sci:99} where the occupied
conformational space of a protein abruptly reduces when approaching
the native state.

Another difference arises in their thermodynamic and structural
behaviors. As shown in Fig.~\ref{fig:500mers} the collapse transition
for the protein-like heteropolymer is significantly shifted towards
higher $T$ as compared to the random sequence ($T_c \approx 1.2$ and
$T_c \approx 0.9$, respectively). This can be rationalized by the
``design'' of the protein-like sequences favoring the formation of
hydrophobic cores already at $T_c$, with a significantly lower energy
compared to the random case (see the corresponding structures with
energies $-240$ and $-193$, respectively). Interestingly, however,
among sequences of the same \emph{type} (\ie either random or
protein-like), $T_c$ shows only little variation. For both sequences,
the drop-off of $R_g$ coincides generally better with the upper
specific heat peak (collapse transition) as compared to the shorter HP
sequences studied above.

The specific heat of the protein-like sequence shows a clear
separation between collapse and folding transition (although the
latter is often manifested by a shoulder only) exemplifying a rough
but structured energy landscape. The course of the winding measure
($\tau$) corroborates this observation signaling a division between
collapse and folding phase. In contrast, for the random heteropolymer,
the two transitions are more interleaved, and the collapse transition
is largely suppressed by the generally stronger transition to the
ground state. This is a consequence of the inability to form a
hydrophobic core and the absence of conformational ``guidance''
(folding path) towards the ground state. Although $\tau$ features also
a more or less pronounced drop-off, its occurrence does not coincide
with the low $T$ peak in the specific heat anymore (as observed for
some of the shorter protein-like HP sequences above).

A systematic analysis of these observations, by considering a large
ensemble of HP sequences, would be both a computational challenge and
an interesting topic to better understand the generic behavior of
these model proteins.

\section{\label{sec:conclusion}Conclusion}

We have presented a generic Monte Carlo scheme based on Wang-Landau
sampling with suitable trial moves (combination of pull,
bond-rebridging and pivot moves) which is very efficient for the
simulation of HP lattice proteins and the analysis of their
thermodynamic properties over the entire temperature range. Because of
its intrinsic simplicity and flexibility, our method is ideally suited
for the many variations of this common protein model, \eg protein
adsorption,\cite{li:comput_phys_commun:11, li:12} protein aggregation
and the like. For instance, with our present, optimized, algorithm, it
takes only about 5.5 hours ($t_{\text{DOS}}$) to obtain the entire DOS
of an HP 36mer adsorbing on a weakly attracting surface (\ie between
one to two orders of magnitude faster than in
Ref.~\onlinecite{li:comput_phys_commun:11}). This system has been used
as a benchmark to demonstrate the efficiency of a recently proposed
Wang-Landau scheme coupled with configurational bias Monte Carlo which
achieved $t_{\text{DOS}} \approx 28$h under the same
conditions.\cite{radhakrishna:j_chem_phys:12} Besides considerable
higher performance, the main advantage of our approach is that its
efficiency does not depend on the tuning of external simulation
parameters (\eg an unphysical temperature as in
Ref.~\onlinecite{radhakrishna:j_chem_phys:12}). Our approach has also
proven powerful for exploring the low-temperature thermodynamics of
interacting self-avoiding walks, even for chain lengths $N \gg
1000$.\cite{wuest:phys_rev_lett:09, wuest:isaw}

Among the various ingredients used to tackle simulations of lattice
proteins or polymers at low temperatures and high densities, ranging
from the Monte Carlo driver to various tuning schemes, it turned out
that the use of appropriate Monte Carlo trial moves has the greatest
impact on the overall performance and accuracy of the
algorithm. However, we stress that it is the interplay between
Wang-Landau sampling, trial move set and efficient implementation
which ultimately resulted in the overall robustness and performance of
our approach.

Substantial further efficiency improvements would likely be gained
with suitable parallelization techniques. Subdivision of the sampling
of the DOS into smaller energy windows or the use of multiple random
walkers which simultaneously update a single DOS are two already well
studied approaches in this direction.\cite{zhan:comput_phys_commun:08}
A more advanced and interesting parallelization scheme would allow
exchange of conformations among several \emph{independently} running
Wang-Landau samplers, similar to the ideas employed in parallel
tempering.\cite{landau:09, *newman:99} Such ``conformational mixing''
could considerably reduce the tunneling time of a random walk between
the energy boundaries and thus speed up the overall convergence of the
simulation.

\begin{acknowledgments}

  We would like to thank Claire Gervais for many interesting and
  fruitful discussions and careful reading of the manuscript. In part,
  this work was supported by NSF Grant No. DMR-0810223.

\end{acknowledgments}

\end{document}